# Ultrahigh refractive index sensitivity of TE-polarized electromagnetic waves in graphene at the interface between two dielectric media

O.V. Kotov,[1,2] M.A. Kol'chenko,[3] and Yu. E. Lozovik[1,4,*]

[1]*Institute for Spectroscopy, Russian Academy of Sciences,142190, Troitsk, Moscow, Russia*
[2]*MIEM at National Research University HSE, 109028, Moscow, Russia*
[3]*Samsung Advanced Institute of Technology, 127018, Moscow, Russia*
[4]*Moscow Institute of Physics and Technology, 141700, Dolgoprudny, Moscow Region, Russia*
[*]*lozovik@mail.ru*

The behavior of the TE and TM electromagnetic waves in graphene at the interface between two semi-infinite dielectric media is studied. The dramatic influence on the TE waves propagation even at very small changes in the optical contrast between the two dielectric media is predicted. Frequencies of the TE waves are found to lie only in the window determined by the contrast. We consider this effect in connection with the design of graphene-based optical gas sensor. Near the frequency, where the imaginary part of the conductivity of graphene becomes zero, ultrahigh refractive index sensitivity and very low detection limit are revealed. The considered graphene-based optical gas sensor outperforms characteristics of modern volume refractive index sensors by several orders of magnitude.

## I. INTRODUCTION

The field of plasmonics [1–7] attracts a great attention of researchers due to the variety of novel phenomena and applications. In particular, plasmonics is an essential component for the design of most metamaterials [8–14]. Over the last few years, the palm of supremacy in plasmonics has been captured by graphene - a two-dimensional (2D) layer of carbon atoms arranged in a honeycomb lattice, which possesses record high carrier mobility [15–17]. Nowadays, graphene plasmonics [18, 19] is a rapidly growing area of physics which causes an enormous interest not only due to the unique properties of intrinsic graphene plasmons [20–37], but also to the optical properties of graphene-based hybrid plasmonic structures [38–56]. Graphene plasmons have several advantages over plasmons in thin metal layers. Despite the same energy they have less decay length due to shorter wavelength and, hence, higher plasmon confinement [28, 30]. Moreover, due to higher carrier mobility in graphene they possess longer propagation distances. But the most important advantage is the capability to dynamically tune the conductivity of graphene by means of chemical doping or gate voltage [40]. All these theoretically predicted advantages have recently been proved by the near-field optical microscopy experiments [57, 58]. The prospects of novel photonic and optoelectronic applications of graphene can be also connected with the existence of a transverse (TE) electromagnetic mode in monolayer [59] and bilayer [60] graphene, as a consequence of their double band electron structure. The dispersion of TE waves is very close to the light line which leads to their very small field confinement. However, as we will show, TE waves possess ultrahigh sensitivity to the changes in the optical contrast between the two semi-infinite dielectric media, located on the opposite sides of graphene.

The dispersion relation $\omega(q)$ of electromagnetic waves with TE (transverse electric) and TM (transverse magnetic) polarization, propagating in the 2D electron system, surrounded by the homogeneous dielectric medium with permittivity $\varepsilon$, and decaying exponentially in the transverse directions, is given by [61, 62]:

$$\frac{\varepsilon}{\sqrt{q^2 - \varepsilon(\omega/c)^2}} = \frac{2\pi\sigma(\omega)}{i\omega} \quad \text{(TM waves)}, \qquad (1)$$

$$\sqrt{q^2 - \varepsilon(\omega/c)^2} = -\frac{2\pi\sigma(\omega)\omega}{ic^2} \quad \text{(TE waves)}, \qquad (2)$$

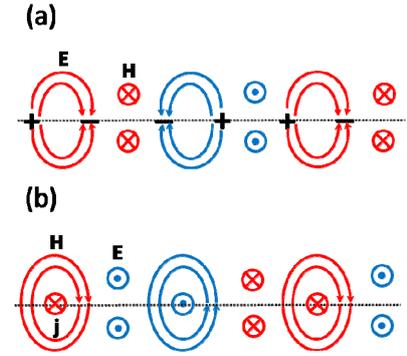

Fig. 1. Schematic representation of TM (a) and TE (b) waves in 2D electron system (e.g., graphene layer) depicted by dotted line. (a) The charge density oscillations for TM waves can be represented in terms of electric dipole wave. (b) Self-sustained oscillations of the current in the case of TE waves can be described in terms of magnetic dipole wave where electric field is always directed opposite to the current.

where $\sigma(\omega)$ is the local dynamic conductivity of the 2D electron system and $c$ is the velocity of light in the free space. The dispersion relation (1) for TM waves (also known as surface plasmon polaritons) in the nonretarded limit ($q \gg \omega/c\sqrt{\varepsilon}$) reduced to the plasmon dispersion. In fact, collective oscillations of 2D charge density $\rho$ described by 2D plasmons are excited by the in-plane electric field $\vec{E}_\parallel$ of incident light: $\partial\rho/\partial t + div\vec{j} = 0$, where $\vec{j} = \sigma(\omega)\vec{E}_\parallel\delta(z)$ is the in-plane electric current as a $\delta$-response of 2D electron system to the $\vec{E}_\parallel$. The resulting pattern of the charge density oscillations for TM waves can be represented in terms of electric dipole wave (see Fig. 1(a)). However, TE waves cannot be reduced to the common plasmons. Since their in-plane electric field oscillations are perpendicular to the propagation vector $q$ ($div\vec{E}_\parallel = 4\pi\rho/\varepsilon = 0$), the electric current is also perpendicular to $q$ ($div\vec{j} = 0$) and 2D charge density $\rho$ is zero. The resulting pattern of self-sustained oscillations of the current in the case of TE waves can be described in terms of magnetic dipole wave. Figure 1(b) shows schematic representation of this wave: induced currents provide local magnetic dipoles with corresponding magnetic field; electric field is always directed opposite to the current (that follows from the condition of TE wave existence $\text{Im}\,\sigma(\omega) < 0$). Actually, as seen from Eqs. (1) and (2) TM waves may exist if $\text{Im}\,\sigma(\omega) > 0$ and TE waves if $\text{Im}\,\sigma(\omega) < 0$. Hence, TE waves cannot exist in conventional 2D electron systems where conductivity can be described by the Drude model which implies $\text{Im}\,\sigma(\omega) > 0$.

The previous works related to the study of TE waves in graphene [27, 31, 41, 59, 60, 63–70] are mainly focused on the case when graphene sheet is embedded into a homogeneous medium or devoted to the investigation of quasi-TE waves. The case of different dielectrics above and below graphene, though mentioned in Refs [60, 70], was not under a detailed consideration. In our work we make consistent calculations of the behavior of TE and TM waves in graphene at the interface between two semi-infinite dielectric media. We show that unlike TM waves, the behavior of TE waves strongly depends on the small changes in the optical contrast between the two dielectric media. We argue that TE waves do not exist in some frequency range depending on the contrast even at $\text{Im}\,\sigma(\omega) < 0$. Solving the electrostatic problem it is easy to show that optical contrast has no dramatic influence on the common plasmon dispersion where dielectric constants of surrounding media are included as the half-sum. We obtain that the same situation will be for plasmon polaritons (TM waves). Here for the first time we estimate TE waves refractive index sensitivity and detection limit in connection with the design of graphene-based optical gas sensor. We propose a novel approach for volume optical sensing employing surface TE waves (STE) in graphene which incorporates some features of the surface plasmon resonance (SPR) sensing [71] and volume optical sensing [72, 73].

## II. GRAPHENE BETWEEN TWO DIELECTRIC MEDIA

Let us consider graphene at the interface between two semi-infinite dielectric media. Usually (see [71]), the sensitivity to the changes in the optical contrast is expressed in terms of refractive index. Hereinafter we will operate with normalized quantities: $Q = q/k_F$ and $\Omega = \hbar\omega/E_F$ are normalized longitudinal wave vector and frequency to the Fermi momentum and the Fermi energy, respectively; $K_x = \dfrac{Q}{\Omega}\dfrac{c}{v_F}$ and

$K_{1,2z} = \dfrac{k_{1,2z}}{\Omega}\dfrac{c}{v_f} = \sqrt{K_x^2 - n_{1,2}^2}$ are normalized longitudinal and transverse wave vectors to the free light momentum, respectively. Here $k_{1,2z} = \dfrac{1}{k_F}\sqrt{q^2 - n_{1,2}^2(\omega/c)^2}$ is transverse wave vector normalized to the Fermi momentum and $v_F \approx 10^6$ m/s $\approx c/300$ is the Fermi velocity of electrons in graphene. In the case of different dielectric media above (with refractive index $n_1$) and below (with refractive index $n_2$) graphene layer the Eqs. (1) and (2) can be rewritten as:

$$\frac{n_1^2}{\sqrt{K_x^2 - n_1^2}} + \frac{n_2^2}{\sqrt{K_x^2 - n_2^2}} = -f \quad \text{(TM waves)}, \qquad (3)$$

$$\sqrt{K_x^2 - n_1^2} + \sqrt{K_x^2 - n_2^2} = f \quad \text{(TE waves)}, \qquad (4)$$

where $f = -\dfrac{4\pi\sigma(\Omega)}{ic}$ with $\sigma(\Omega)$ is the local dynamic conductivity of graphene. The dynamic conductivity of graphene in units of $\sigma_0 = e^2/4\hbar$ calculated in random-phase approximation in the local-response limit ($Q \ll 1$) can be expressed as [74, 75]:

$$\frac{\sigma(\Omega)}{\sigma_0} = \Theta(\Omega - 2) + \frac{i}{\pi}\left(\frac{4}{(\Omega + i\Gamma)} - \ln\left|\frac{\Omega + i\Gamma + 2}{\Omega + i\Gamma - 2}\right|\right) \qquad (5)$$

at zero temperature and

$$\frac{\sigma(\Omega,t)}{\sigma_0} = \frac{1}{2} + \frac{1}{\pi}\arctan\left(\frac{\Omega - 2}{2t}\right) + \\ + \frac{i}{\pi}\left(\frac{8t\ln\left(2\cosh\left(\dfrac{1}{2t}\right)\right)}{(\Omega + i\Gamma)} - \frac{1}{2}\ln\left(\frac{(\Omega + i\Gamma + 2)^2}{(\Omega + i\Gamma - 2)^2 + (2t)^2}\right)\right) \qquad (6)$$

at finite temperatures, where $t = T/E_F$, $\Gamma = \hbar\tau^{-1}/E_F$, with $T$ and $\tau$ are temperature in units of energy and a finite carrier relaxation time in graphene, respectively. Imaginary part of conductivity becomes zero at $\Omega_0 \approx 1.667$ for $T = 0K$ and $\Omega_T \approx 1.67$ for $T = 300K$ (see Fig. 2).

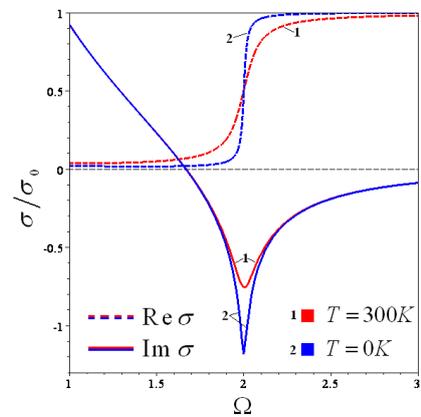

Fig. 2. The real and imaginary parts of the dynamic conductivity of graphene, in units of $\sigma_0 = e^2/4\hbar$ as a function of frequency $\Omega = \hbar\omega/E_F$ at zero and room temperatures. The parameters of graphene are set as $E_F = 1\text{eV}$, $\tau = 0.5 \cdot 10^{-13}\text{s}$.

That is TE waves exist only at $\Omega > 1.667 (\Omega > 1.67)$ for $T = 0K (300K)$. Due to the Landau damping at $\Omega > 2$ (at finite temperatures a little less) TE waves should be excited only in the range $1.667 < \Omega < 2$. To reduce the role of the temperature here and below we will consider high doped graphene: $E_F = 1\text{eV}$. For high frequencies (above the phonon frequency $\sim 0.2 eV$) the carrier relaxation time in graphene is mainly determined by electron-phonon scattering mechanism and can be incorporated through an effective $\tau = 0.5 \cdot 10^{-13}\text{s}$ (mobility $\mu = 10^4 \text{cm}^2/\text{Vs}$) [28]. In case of TE waves it is easy to find complex analytical solution of the Eq. (4) by solving the system of complex equations:

$$\begin{cases} K_{1z} + K_{2z} = f, \\ K_{1z}^2 + n_1^2 = K_{2z}^2 + n_2^2. \end{cases} \qquad (7)$$

Solving system (7) we obtain the real and imaginary parts of normalized transverse wave vectors:

$$K_{1z} = \frac{\text{Re}\,f\left(|f|^2 + (n_2^2 - n_1^2)\right)}{2|f|^2} + i\frac{\text{Im}\,f\left(|f|^2 - (n_2^2 - n_1^2)\right)}{2|f|^2}, \qquad (8)$$

$$K_{2z} = \frac{\text{Re}\,f\left(|f|^2 - (n_2^2 - n_1^2)\right)}{2|f|^2} + i\frac{\text{Im}\,f\left(|f|^2 + (n_2^2 - n_1^2)\right)}{2|f|^2}. \qquad (9)$$

Setting $n_2 > n_1$, we obtain that for $(n_2^2 - n_1^2) > |f(\Omega)|^2$ (see Eq. (9)) $\operatorname{Re} K_{2z}$ becomes negative, which leads to an exponential growth of the wave field with the distance from graphene into the medium with refractive index $n_2$. Such solutions should be rejected as unphysical, because they do not satisfy the boundary conditions at infinity. Therefore, when the relative permittivity of dielectrics above and below graphene differs more than the $|f(\Omega)|^2$, TE wave cannot propagate along graphene layer. Further we will consider this effect applied to the design of graphene-based optical gas sensor. Its possible registration system is shown in Fig. 3.

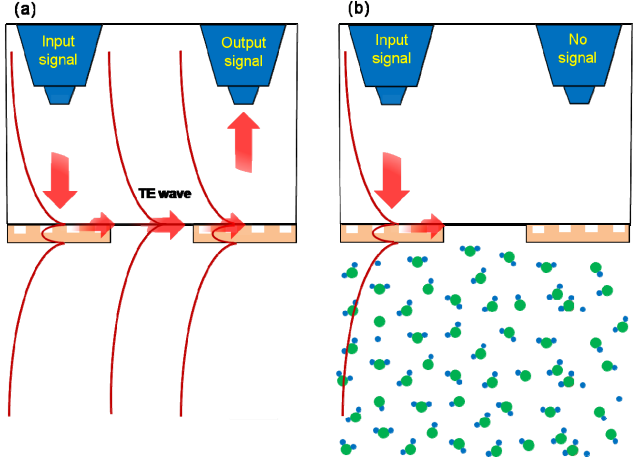

Fig. 3. The possible registration system of graphene-based optical gas sensor (see the text). (a) Graphene is surrounded by media with equal refractive index $n_1 = n_2 = 1$. (b) After the appearance of the investigated gas refractive index below the graphene layer is increased by $n_x$ (i.e. $n_1 = 1$ and $n_2 = 1 + n_x$).

Incident light excites TE wave in graphene, e.g., by means of grating substrate, then, after passing the suspended part of graphene, TE wave decouples to light by another grating substrate (see Fig. 3(a)). We assume that after the appearance of the investigated gas refractive index below the graphene layer ($n_2$) is increased by $n_x$ while refractive index above the graphene layer ($n_1$) remains the same (i.e. $n_1 = 1$ and $n_2 = 1 + n_x$). Thus, the condition of TE wave nonexistence $(n_2^2 - n_1^2) = n_x^2 + 2n_x > |f(\Omega)|^2$ mentioned above is determined by the refractive index change $\Delta n = n_x$ as a function of frequency:

$$n_x(\Omega) = \sqrt{1 + |f(\Omega)|^2} - 1 \approx |f(\Omega)|^2 / 2. \quad (10)$$

If the concentration of the investigated gas exceeds critical value corresponding to $n_x$, TE wave will no longer exist in suspended part of graphene and, ideally, there will be no output signal (see Fig. 3(b)). Using Eqs. (5) and (6) we plot the function $n_x(\Omega) = \frac{1}{2}\left|\frac{4\pi\sigma(\Omega,t)}{ic}\right|^2$ at zero and room temperatures (Fig. 4(a)). The inset from Fig. 4(a) shows that the smallest detectable refractive index change $(n_x)_{\min}$ (minimal detection limit) at zero and room temperatures differs by several orders of magnitude. It is determined by the Eq. (10) where it is set that $\Omega = \Omega_0$. Due to the identity $\operatorname{Im}\sigma(\Omega = \Omega_0) \equiv 0$ we have: $(n_x)_{\min} = \sqrt{1 + |\operatorname{Im} f(\Omega_0)|^2} - 1$, where

$|\operatorname{Im} f(\Omega_0)|^2 = \left|\frac{4\pi}{c}\operatorname{Re}\sigma\right|^2 \propto \left(\frac{2T\ln(2\cosh(1/2T))}{\tau}\right)^2$. At room temperature $(n_x)_{\min}$ does not change with the increasing of carrier relaxation time in graphene and equals $(n_x)_{\min} = 6.7 \cdot 10^{-7}$ RIU, where RIU stands for refractive index units. While at zero temperature $|\operatorname{Im} f(\Omega_0)|^2 \propto \left(\frac{1}{\tau}\right)^2$ and, hence, $(n_x)_{\min}$ decreases by two orders of magnitude with the increase of mobility by one order of magnitude (mobility in high-quality suspended graphene can be improved even up to $\mu = 2 \cdot 10^5 \, \text{cm}^2/\text{Vs}$ [17]). At $\tau = 0.5 \cdot 10^{-13}$ s ($\mu = 10^4 \, \text{cm}^2/\text{Vs}$) we get $(n_x)_{\min} = 3.2 \cdot 10^{-12}$ RIU and at $\tau = 0.5 \cdot 10^{-12}$ s ($\mu = 10^5 \, \text{cm}^2/\text{Vs}$) we obtain $(n_x)_{\min} = 3 \cdot 10^{-14}$ RIU (see Fig. 4(b)).

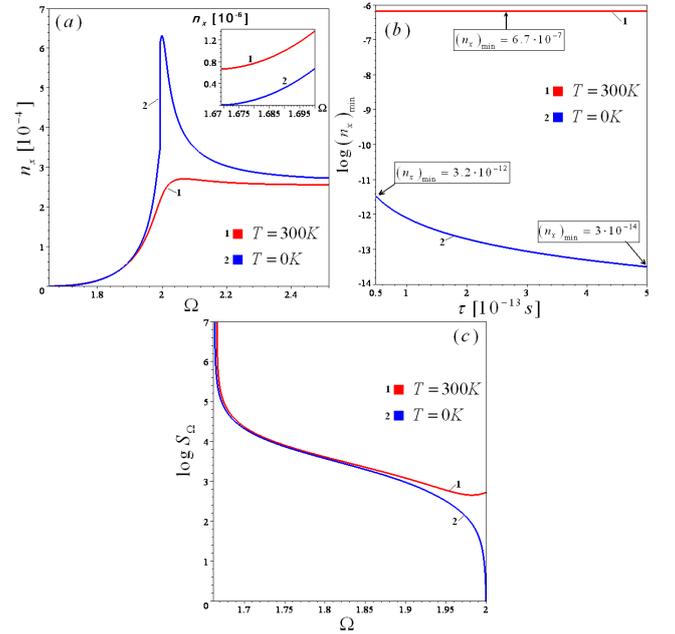

Fig. 4. The function of refractive index change $n_x(\Omega)$ (inset: the same near $\Omega = \Omega_0$) (a), common logarithm of the minimal detection limit as a function of carrier relaxation time (b) and common logarithm of the refractive index sensitivity in units RIU$^{-1}$ as a function of frequency $\Omega = \hbar\omega/E_F$ (c) at zero and room temperatures. The parameters of graphene are set as $E_F = 1\text{eV}$, $\tau = 0.5 \cdot 10^{-13}$ s (for (a) and (c)).

The refractive index sensitivity (sensitivity to the refractive index changes) depending on the normalized frequency can be written as: $S_\Omega = \frac{\Delta\Omega}{\Delta n_x} = \left(\frac{dn_x(\Omega)}{d\Omega}\right)^{-1}$ [RIU$^{-1}$]. Figure 4(c) shows that near $\Omega = \Omega_0$ the sensitivity do not depend on temperature and tends to infinity. Further we will refer to the frequency at which the sensitivity reaches its maximum value as the sensitivity point. But in fact it is limited by the value corresponding to the required confinement (see below) and by the accuracy of setting the necessary frequency of the wave. The last is defined by the charge inhomogeneity in graphene [76] which leads to the accuracy of the Fermi level $\Delta E_F \sim 10^{-3}$ eV [77] and by the accuracy of the thermal control which is typically $\Delta T \sim 10^{-3} K$. Using Eq. (6) at $E_F = 1\text{eV}$ and $\tau = 0.5 \cdot 10^{-13}$ s we get $\Delta\Omega_0 \sim 10^{-6}$ for the charge inhomogeneity factor and $\Delta\Omega_0 \sim 10^{-7}$ for the temperature fluctuations. The dispersion relation of TE waves $\Omega(Q)$ defined by the Eq. (4) can be easily found from the complex equation $K_x^2 = \left(\dfrac{Q}{\Omega}\dfrac{c}{v_F}\right)^2 = K_{1z}^2 + n_1^2 \equiv F(K_{1z})$:

$$Q(\Omega) = \Omega \frac{v_F}{c}\left(\sqrt{\frac{\operatorname{Re} F(K_{1z}) + |F(K_{1z})|}{2}} + i\frac{\operatorname{Im} F(K_{1z})}{\sqrt{2(\operatorname{Re} F(K_{1z}) + |F(K_{1z})|)}}\right), \quad (11)$$

where $F(K_{1z}) = (\operatorname{Re} K_{1z} + i\operatorname{Im} K_{1z})^2 + n_1^2$ and $K_{1z}$ defined by the Eq. (8). In the case, when the damping can be neglected, $\operatorname{Im} f = 0$ and Eq. (11) takes the simple form:

$$Q(\Omega) = \Omega \frac{v_F}{c}\left(\sqrt{\frac{(f^2 + n_2^2 - n_1^2)^2}{4f^2} + n_1^2}\right). \quad \text{For} \quad n_1 = n_2$$

$Q(\Omega) = \Omega\dfrac{v_F}{c}\left(\sqrt{(f/2)^2 + n_1^2}\right)$, i.e. at $\Omega = \Omega_0$ the dispersion of TE waves goes to the dispersion of light in the medium surrounding graphene: $Q(\Omega) = \Omega\dfrac{v_F}{c}n_1$. For $n_1 \neq n_2$ at the frequency $\Omega_n$ corresponding to the condition of TE wave nonexistence $|f(\Omega_n)|^2 = n_2^2 - n_1^2$ (see Eq. (10)) the dispersion of TE waves goes to the dispersion of light in the medium with the highest refractive index: $Q(\Omega) = \Omega\dfrac{v_F}{c}n_2$. It always lies to the right of the most inclined light line and, hence, cannot exist as leaky modes. Taking into account the damping, the dispersion relation of TE waves, which is generally defined by the Eq. (11), is also very close to the dispersion of light. At $T = 300K$ for $n_x = 10^{-5}$ RIU TE waves exist at $\Omega > 1.78$ and for $n_x = 10^{-6}$ RIU at $\Omega > 1.69$ (see Fig. 5(a)), which is in agreement with the condition (9). The Eq. (3) for TM waves has to be solved numerically. As we have expected the optical contrast has no dramatic influence on the TM waves dispersion (see Fig. 5(b)).

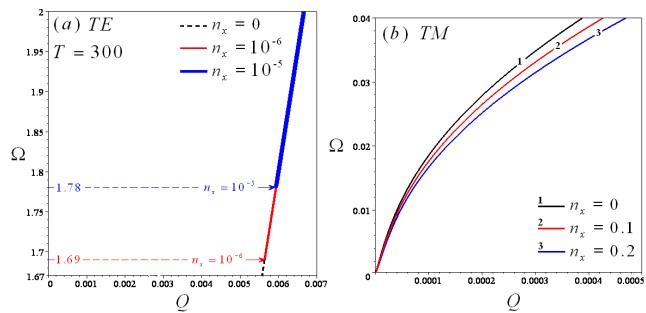

Fig. 5. The dispersion of TE (a) and TM (b) waves. (a) For $n_x = 0$ (black line), $n_x = 10^{-6}$ RIU (red line), $n_x = 10^{-5}$ RIU (blue line). (b) For $n_x = 0$ (black line), $n_x = 0.1$ RIU (red line), $n_x = 0.2$ RIU (blue line). The parameters of graphene are set as $E_F = 1\text{eV}$, $\tau = 0.5\cdot 10^{-13}\text{s}$.

### III. TE WAVES CONFINEMENT

Let us consider in more detail what happens to the wave vectors with the increasing of the optical contrast. The normalized transverse wave vectors $K_{1,2z} = \dfrac{k_{1,2z}}{\Omega}\dfrac{c}{v_f}$ introduced above express the degree of wave confinement. Indeed, wave confinement can be taken as $\lambda/(2\pi L_z)$, where $\lambda$ is the wavelength in air and $L_z = 1/\operatorname{Re}(k_z)$ is the wave decay length in the transverse direction corresponding to the $1/e$ field decay. Then we get: $\dfrac{\lambda}{2\pi L_z} = \dfrac{\lambda}{2\pi}\operatorname{Re}(k_z) = \dfrac{c}{\omega}\dfrac{\Omega v_f}{c}\operatorname{Re}(K_z)k_f = \operatorname{Re}(K_z)$.

For the case of common plasmons in the nonretarded limit $\operatorname{Re}(k_z) = 2\pi/\lambda_{pl}$ and the expression for confinement takes the usual form $\lambda/\lambda_{pl}$. For zero refractive index change ($n_x = 0$) $K_{1z} = K_{2z}$ defined by the Eqs. (8) and (9) shown in Fig. 6(a). At frequencies $\Omega < 1.667 (\Omega < 1.67)$ for $T = 0K(300K)$ the real part of transverse wave vectors $\operatorname{Re} K_{1,2z}$ get negative and TE wave does not exist. From Fig. 6(a) it is seen that TE wave confinement in graphene of the order of magnitude $10^{-2}$ while for common plasmons in graphene it can reach values of the order of magnitude $10^2$. On the appearing of the refractive index change less than the smallest detectable one (e.g. $n_x = 6.6 \cdot 10^{-7}$ RIU $< (n_x)_{\min}$) the TE wave confinement at the side with the highest refractive index ($\operatorname{Re} K_{2z}$) decreases and at the opposite side ($\operatorname{Re} K_{1z}$) increases (see Fig. 6(b) green line (2)). When the refractive index change begins to exceed $(n_x)_{\min}$ (e.g., $n_x = 10^{-5}$ RIU) the behavior of the TE wave confinement changes significantly (see Fig. 6(b) red line (3)). The confinement $\operatorname{Re} K_{1z}$ continues to increase and $\operatorname{Re} K_{2z}$ continues to decrease, but in such a way that at the frequency $\Omega = 1.78$ it becomes zero. This is in agreement with the results for TE waves dispersion represented by the Fig. 5(a). Thus, the absence of TE waves at frequencies less that those which satisfy the Eq. (10) is caused by the delocalization of the wave at the side with the highest refractive index (the side filled with the investigated gas). In the sensitivity point (where the sensitivity $S$ goes to infinity) the confinement $\operatorname{Re} K_{2z}$ goes to zero. With the increase of $n_x$ the sensitivity point shifts towards the damping region (see Fig. 6(b)).

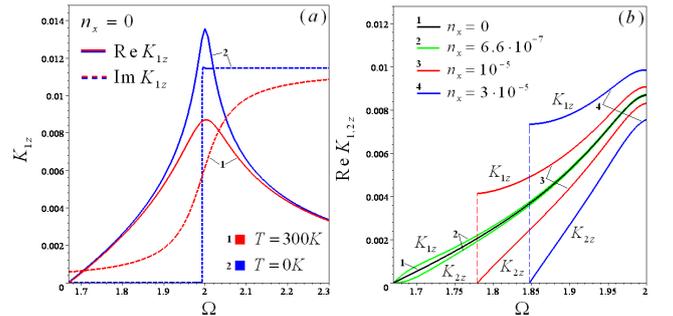

Fig. 6. The normalized transverse wave vectors $K_{1,2z}$ of TE waves in graphene as a function of frequency. (a) Real and imaginary parts of $K_{1,2z}$ for $n_x = 0$ at zero and room temperatures. (b) Real part of $K_{1,2z}$ (expressing wave confinement) at $T = 300K$ for: $n_x = 0$ (black (1)), $n_x = 6.6 \cdot 10^{-7}$ RIU (green (2)), $n_x = 10^{-5}$ RIU (red (3)) and $n_x = 3 \cdot 10^{-5}$ RIU (blue (4)). The parameters of graphene are set as $E_F = 1\text{eV}$, $\tau = 0.5 \cdot 10^{-13}\text{s}$.

The above calculations were carried out under the assumption that graphene is surrounded by two semi-infinite media. In reality one deals with finite volume filled by the investigated gas. In order that the upper and lower boundaries of the medium above and below the graphene layer, respectively, do not affect the TE wave refractive index sensitivity, the wave decay length $L_z$ (see above) should be less than the required distances above and below the graphene layer. In this case the wave will not bound with upper or lower boundaries, and so their influence can be neglected. The quantity $L_z = 1/|k_z|$ determines the $1/e$ field decay, where $k_z = \sqrt{Q^2 - n_{1,2}^2 (\Omega v_F/c)^2}$. For the proper comparison of our results with the characteristics of modern refractive index sensors let us express the decay length and the sensitivity depending on the wavelength in non-normalized variables: $L_z$ in the units of length and $S_\lambda = \frac{\Delta \lambda}{\Delta n_x} = \left(\frac{dn_x(\lambda)}{d\lambda}\right)^{-1}$ in nm/RIU. The sensitivity reaches its maximum value near the point $\Omega_0 \approx 1.667$ ($\Omega_T \approx 1.67$) (see Fig. 4(c)) which corresponds to $\lambda_0 \approx 744$nm ($\lambda_T \approx 743$nm) at $E_F = 1$eV. On the other hand, the decay length defining the transverse size of the investigated volume grows with the increase of the wavelength. Depending on the required measurements one should find the optimum balance between the decay length and the sensitivity. Let us consider the dependences $L_z(\lambda)$ at $n_x = 0$ and $\log S_\lambda(\lambda)$ on the same plot. At both temperatures $T = 0K$ and $T = 300K$ for the required transverse size larger than $L_{2z} = 1$mm the sensitivity will be $S_\lambda \simeq 5 \cdot 10^7$ nm/RIU (see Fig. 7(a)). For the corresponding wavelengths $\lambda = 740.81$nm ($\lambda = 739.63$nm) at $T = 0K$ ($T = 300K$) the detection limit defined by the Eq. (10) is $n_x = 3 \cdot 10^{-8}$ RIU ($n_x = 7.3 \cdot 10^{-7}$ RIU). Figure 7(b) shows the decay length and the sensitivity at wavelengths near the damping region (corresponds to $\Omega \to 2$). At both temperatures we find $S_\lambda \sim 10^6$ nm/RIU. At $\lambda = 685.2$nm the decay length and the detection limit will be $L_z = 40\mu$m and $n_x = 1.7 \cdot 10^{-5}$ RIU, respectively. From Fig. 7(b) one can see that at $\lambda = 630$nm the minimal transverse size of the investigated volume in our consideration will be $L_z \simeq 10\mu$m.

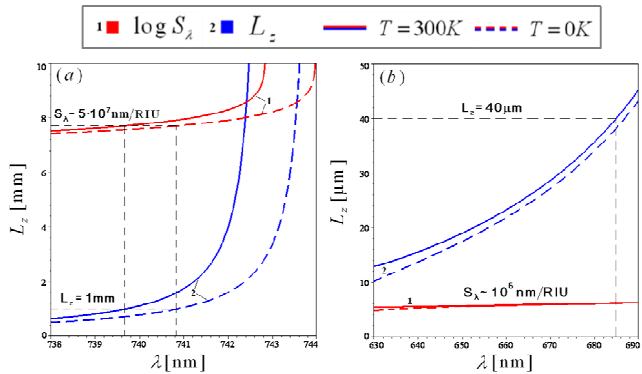

Fig. 7. TE wave decay length in graphene $L_z$ (blue lines) at $n_x = 0$ and common logarithm of the refractive index sensitivity $S_\lambda$ [nm/RIU] (red lines) as a function of wavelength at $T = 0K$ (dashed lines) and at $T = 300K$ (solid lines) for wavelengths near the sensitivity point (a) and for wavelengths near the damping region (b). The parameters of graphene are set as $E_F = 1$eV, $\tau = 0.5 \cdot 10^{-13} s$.

**IV. DISCUSSION**

Let us discuss some details which can be important in experiment. Our calculations were held for high Fermi level of graphene ($E_F = 1$eV) in order to reduce the role of the temperature. Such Fermi level can be achieved in graphene only by means of strong chemical doping, which will lead to a very high suppression of the carrier mobility in graphene. However, as can be seen from Fig. 4(b) at $T = 300K$ the refractive index resolution is almost independent from the carrier mobility and, hence, strong doping does not impair the performance of the sensor. At low temperatures (for helium temperature all results will be almost the same as for $T = 0K$) it is not necessary to use high doped graphene. It is possible to work at low Fermi levels (less than $0.2eV$) and to achieve very high carrier mobility in graphene, which can improve the refractive index resolution by several orders of magnitude (see Fig. 4(b)).

Some difficulties may occur in searching the optimal length of the suspended part of graphene in the direction of the wave propagation. On the one hand, in order to avoid a mechanical sagging of graphene the length should be several or a few tens of microns. In this case, the registration system suggested here (see Fig. 3) may give the suppression of the output signal depending on the length of the suspended part of graphene rather than its absence. Possibly, for sufficiently small suspended graphene length the considered effect can become unobservable. On the other hand, at small distances between left and right grating couplers it will be difficult to provide low-background measurement with independent illumination of the left coupler and collection signal light from the right coupler. The possible solution to this is placing the coupler and decoupler far enough from the suspended part of graphene. But the distance between each of gratings and the suspended part should not exceed the propagation length of TE waves. Since the propagation length is of the order of several hundred microns it seems possible to distance the coupler and decoupler by considerable measure, while the length of the suspended part of graphene remains several microns.

The main problem of the experimental observation of TE waves in graphene is their very small field confinement. It can be not so easy to distinguish TE wave propagating along the graphene layer from the total electromagnetic field of incident light. In fact, there are different ways to improve TE waves confinement in graphene. Combining graphene with waveguide [41, 78] or making multilayer graphene system [63, 79] it is possible to get high-confined quasi-TE waves containing waveguide component. Perhaps, it will be possible to increase the confinement by the usage of strained graphene sheets [64], or by the applying of quantizing magnetic field leading to the hybrid TM-TE waves [66-68]. Also, TE wave confinement may become higher if we take into account the spatial dispersion effect in graphene [69]. But the most possible solution to the problem can be the usage of another atom-thick systems with richer electron band structure instead of monolayer graphene. An example of such a system can be a Bernal-type bilayer graphene where TE wave confinement can be improved in comparison with monolayer graphene by two orders of magnitude [60]. However, it is important to emphasize that TE wave confinement described by $\text{Re}(K_{1,2z})$ is proportional to the absolute value of $\text{Im}\sigma(\omega)$ (see Eqs. (8), (9)), whereas their sensitivity, discussed here, is *inversely* proportional to the gradient of $\text{Im}\sigma(\omega)$. Due to the monotonic behavior of $\text{Im}\sigma(\omega)$ near the sensitivity point, the increase of the absolute value, which causes the improvement of the confinement, inevitably leads to the increase of the gradient, which results in the decrease of sensitivity. In other words high refractive index sensitivity of TE waves in graphene, predicted in this work, is a reverse side of their small field confinement.

In connection with the comparison of our results with the characteristics of modern refractive index sensors it should be noted that STE sensing proposed here incorporates some features of SPR sensing [71] and volume optical sensing [72, 73]. Both SPR and STE sensing are based on the interaction between a sample and an evanescent electromagnetic wave. But SPR sensing uses common plasmons with typically small field confinement (10-300 nm) and, hence, operates with thin layers of analyte. On the other hand, based on the beam deviation technique volume optical sensing similarly to STE sensing operates with tens of micrometers of the investigated gas or liquid. Thus, due to the similarity of application areas, it will be more correct to compare our results with the characteristics of volume optical sensors.

Finally, we would like to mention that our calculations were held under the assumption of the homogeneous density of the investigated gas. Otherwise (particularly when we do not try to investigate gas but very thin layers on the surface of graphene) it is necessary to calculate the scattering of TE waves by a finite-size barrier located on the graphene surface. This problem will be the subject of the future investigation.

## V. CONCLUSION

To conclude, we have shown that, unlike TM electromagnetic waves, TE waves in graphene at the interface between two semi-infinite dielectric media can exist only in some frequency range depending on the optical contrast between these media. We have obtained the analytical expressions describing the TE waves frequency range and their sensitivity to the changes in the optical contrast. The effect was considered in connection with the design of graphene-based optical gas sensor. We have found that near the frequency, where the imaginary part of the conductivity of graphene becomes zero, this sensor may have very high refractive index sensitivity and very low detection limit. At zero temperature we found the *minimal* detection limit to be $(n_x)_{\min} = 3.2 \cdot 10^{-12}\,\mathrm{RIU}$. Moreover, the increase of the carrier mobility in graphene by one order of magnitude at zero temperature leads to the decrease of the *minimal* detection limit by two orders of magnitude. At room temperature the *minimal* detection limit is $(n_x)_{\min} = 6.7 \cdot 10^{-7}\,\mathrm{RIU}$. For the transverse size of the investigated volume larger than $L_z = 1\mathrm{mm}$ we found that the sensitivity is $S_\lambda \simeq 5 \cdot 10^7\,\mathrm{nm/RIU}$ and the detection limit is $n_x = 3 \cdot 10^{-8}\,\mathrm{RIU}$ ($n_x = 7.3 \cdot 10^{-7}\,\mathrm{RIU}$) at $T = 0K$ ($T = 300K$). The minimal operating transverse size of the investigated volume for the considered sensor is $L_z \simeq 10\mu\mathrm{m}$. For $L_z = 40\mu\mathrm{m}$ the sensitivity and the detection limit will be $S_\lambda \sim 10^6\,\mathrm{nm/RIU}$ and $n_x = 1.7 \cdot 10^{-5}\,\mathrm{RIU}$, correspondingly. The sensitivity of the considered graphene-based optical gas sensor exceeds the sensitivity of volume refractive index sensors based on the beam deviation technique [73] ($L_z \simeq 20\mu\mathrm{m}$, $n_x = 1.7 \cdot 10^{-5}\,\mathrm{RIU}$, $S_\lambda \sim 10^3\,\mathrm{nm/RIU}$) by three orders of magnitude. By changing input signal frequency or Fermi level in graphene one can find the optimum balance between required field confinement and refractive index sensitivity. Unlike SPR sensors, TE waves graphene-based sensor poroposed here, as well as any volume optical sensor, is suitable for applications requiring thick surface functionalization or measurements through bigger biological samples, such as living cells.

### Acknowledgments

The authors are grateful to A. A. Sokolik for useful discussions. The work was supported by the Grant of GRO SAIT and by Russian Foundation for Basic Research. Yu. E. L. thanks the Basic Research Program of the National Research University HSE. O. V. K. acknowledges the support given by the Grant of President of Russian Federation MK-5288.2011.2.